\documentclass[preprintnumbers,amsmath,amssymb,twocolumn,nofootinbib]{revtex4}

\usepackage{graphicx}
\usepackage[font={scriptsize,singlespacing}]{caption}
\usepackage{epstopdf}
\usepackage{amsmath}
\usepackage{amssymb}
\usepackage{subfigure}
\usepackage{hyperref}
\usepackage{url}
\usepackage{xcolor}
\usepackage{color}

\def\sideremark#1{\ifvmode\leavevmode\fi\vadjust{\vbox to0pt{\vss
 \hbox to 0pt{\hskip\hsize\hskip1em
 \vbox{\hsize2cm\tiny\raggedright\pretolerance10000
 \noindent #1\hfill}\hss}\vbox to8pt{\vfil}\vss}}}%
\definecolor{amaranth}{rgb}{0.9, 0.17, 0.31}
\definecolor{purple(munsell)}{rgb}{0.62, 0.0, 0.77}
\definecolor{americanrose}{rgb}{1.0, 0.01, 0.24}
\definecolor{palatinateblue}{rgb}{0.15, 0.23, 0.89}
\definecolor{royalblue(web)}{rgb}{0.25, 0.41, 0.88}
\definecolor{hanpurple}{rgb}{0.32, 0.09, 0.98}
\definecolor{beaublue}{rgb}{0.74, 0.83, 0.9}
\definecolor{carminered}{rgb}{1.0, 0.0, 0.22}
\definecolor{brightpink}{rgb}{1.0, 0.0, 0.5}
\hypersetup{ linktoc=all,
    colorlinks, linkcolor={palatinateblue},
    citecolor={brightpink}, urlcolor={hanpurple}
}

\begin{document}

  \title{Remnant Symmetry, Propagation and Evolution in $f(T)$ Gravity}

	 \author{Pisin Chen}
     \email{pisinchen@phys.ntu.edu.tw}
     \affiliation{(1) Leung Center for Cosmology and Particle Astrophysics \\ \& Graduate Institute of Astrophysics \& Department of Physics,\\  National Taiwan University,
Taipei 10617, Taiwan\\
(2) Kavli Institute for Particle Astrophysics and Cosmology, \\SLAC National Accelerator Laboratory, Stanford University, CA 94305, U.S.A}

\author{Keisuke Izumi}
     \email{izumi@phys.ntu.edu.tw}
     \affiliation{Leung Center for Cosmology and Particle Astrophysics,\\  National Taiwan University,
Taipei 10617, Taiwan} 

\author{James M. Nester}
     \email{nester@phy.ncu.edu.tw}
     \affiliation{(1) Department of Physics \& Graduate Institute of Astronomy\\ \& Center for Mathematics and Theoretical Physics,\\National Central University, Chungli, 320, Taiwan}
		 \affiliation{(2) Leung Center for Cosmology and Particle Astrophysics,\\  National Taiwan University,
Taipei 10617, Taiwan} 
		\altaffiliation[Currently visiting the~]{Morningside Center of Mathematics, 
Academy of Mathematics and System Science, Chinese Academy of Sciences,
55 Zhongguancun Donglu, Haidian District, 
Beijing 100190, China.}

   \author{Yen Chin Ong}
     \email{yenchin.ong@nordita.org}
     \affiliation{Nordita, KTH Royal Institute of Technology and Stockholm University, \\ Roslagstullsbacken 23,
SE-106 91 Stockholm, Sweden}

\begin{abstract}
It was recently argued that $f(T)$ gravity could inherit ``remnant symmetry'' from the full Lorentz group, despite the fact that the theory is not locally Lorentz invariant.
Confusion has arisen regarding the implication of this result for the previous works, which established that $f(T)$ gravity is pathological due to superluminal propagation, local acausality, and non-unique time evolution. We clarify that the existence of  the ``remnant group'' does not rid the theory of these various problems, but instead strongly supports it.
\end{abstract}

\maketitle

\section{Introduction: $f(T)$ Gravity and Remnant Symmetry}

General Relativity [GR] is a geometric theory of gravity formulated on a Lorentzian manifold equipped with the Levi-Civita connection. This connection is torsion-less and metric compatible -- the gravitational field is completely described in terms of the Riemann curvature tensor. However, given a smooth manifold one can equip it with other connections. If one chooses to use the Weitzenb\"ock connection, then the geometry is \emph{flat} -- the connection being curvature-free [but still metric compatible]. The gravitational field is now completely described in terms of the torsion tensor. Surprisingly GR can be recast into ``teleparallel equivalent of GR'' [TEGR, or GR$_{\|}$] which employs the Weitzenb\"ock connection, a subject which has a large set of literature [see, e.g., \cite{Moller:1961jj, Cho:1975dh, Hehl:1994ue, Itin:1999wi, deAndrade:2000kr, BtelepHam, Bgauge, Hehl:2012pi, Pereira.book}].

The dynamical variable of TEGR, as well as its $f(T)$ extension \cite{eric}, is the frame field [vierbein] $\{{e}_a(x)\}$, or equivalently its corresponding co-frame field $\{e^a (x) \}$. 
The vierbein is related to the coordinate vector fields $\{\partial_\mu\}$ by $e_a(x) = e_a^{~\mu} (x) \partial_\mu$, and similarly $e^a(x) =e^a_{~\mu} (x) dx^\mu$.
The vierbein ${{e}_a(x)}$ forms an orthonormal basis
for the tangent space $T_xM$ at each point $x$ of a given spacetime manifold $(M,g)$. 
The metric tensor $g$ is related to the vierbein field by 
\begin{equation} 
g_{\mu\nu}(x)=\eta_{ab}\, e^a_{~\mu} (x)\, e^b_{~\nu} (x).
\end{equation}

The Weitzenb\"ock connection is defined by
\begin{equation}
\overset{\mathbf{w}}{\nabla}_X Y :=(XY^a){e}_a,
\end{equation}
where $Y=Y^a {e}_a$. 
This means that we declare the vierbein field to be \emph{teleparallel}, i.e., covariantly constant: $\overset{\mathbf{w}}\nabla_X {e}_a = 0$. 
Equivalently, the connection coefficients are
\begin{equation} 
\label{Weitzenb}
{\Gamma}^\lambda{}_{
\nu\mu}
=e^\lambda_{~a}\: \partial_\mu
e^a_{~\nu}.
\end{equation}
It is then straightforward to show that this connection is curvature-less but the torsion tensor is nonzero in general.

Under a linear transformation of the bases $\{e_a(x)\}$ of the tangent vector field
\begin{equation}
e_a(x) \to e'_a(x) = L_a^{~b}(x) e_b(x), ~~\text{det}(L_a^{~b}) \neq 0,
\end{equation}
the connection 1-form
\begin{equation}
\Gamma^b{}_a(x) = \langle \theta^b, \overset{\mathbf{w}}{\nabla} e_a \rangle = {\Gamma}^b{}_{\mu a} dx^\mu,
\end{equation}
transforms as 
\begin{equation}
\Gamma'^a{}_{\mu b} = (L^{-1})^a_{~d}\Gamma^d{}_{\mu c}L^c_{~b} + (L^{-1})^a_{~c}L^c_{~b,\mu},
\end{equation}
where the comma in the subscript denotes the usual partial differentiation. 

We would like to emphasize that there is in fact a difference between ``parallelizable'' and ``teleparallel''. A parallelizable manifold $M$ means that there exists a global frame field on $M$ [that is, the frame bundle $FM$ has a global section]. For example, $S^3$ is parallelizable but $S^2$ is not. Whether a manifold is parallelizable or not depends on the topology but \emph{not} on the connection. In 4-dimensions, the necessary and sufficient condition for parallelizability is the vanishing of the second Stiefel-Whitney characteristic class. Teleparallel geometry means one has a \emph{connection} which is flat everywhere, i.e., has vanishing curvature. [A manifold with a teleparallel connection is always parallelizable.]

One then defines the contortion tensor, which is the difference 
between the Weitzenb\"ock and Levi-Civita connections. In component form, it reads
\begin{equation}  
K^{\mu\nu}_{\:\:\:\:\rho}=-\frac{1}{2}\Big(T^{\mu\nu}_{
\:\:\:\:\rho}
-T^{\nu\mu}_{\:\:\:\:\rho}-T_{\rho}^{\:\:\:\:\mu\nu}\Big).
\end{equation}
For convenience, one usually also defines the tensor
\begin{equation}  
S_\rho^{\:\:\:\mu\nu}=\frac{1}{2}\Big(K^{\mu\nu}_{\:\:\:\:\rho}
+\delta^\mu_\rho
\:T^{\alpha\nu}_{\:\:\:\:\alpha}-\delta^\nu_\rho\:
T^{\alpha\mu}_{\:\:\:\:\alpha}\Big).
\end{equation}

In TEGR, the Teleparallel Lagrangian consists of the so-called ``torsion scalar''
\begin{equation}  
T := S_\rho^{\:\:\:\mu\nu}\:T^\rho_{\:\:\:\mu\nu}.
\end{equation}
It turns out that the ``torsion scalar'' only differs from the Ricci scalar [obtained from the usual Levi-Civita connection] by a boundary term: $T = -R + \text{div}(\cdot)$, and so it encodes all the dynamics of GR. One could then promote $T$ to a function $f(T)$, similar to how GR is generalized to $f(R)$ gravity. For a general $f$, this would lead to a dynamical gravity theory that would have second order field equations [whereas $f(R)$ gravity gives higher order equations], with some kind of non-linear dynamics that differs from GR but nevertheless reduces to GR in a certain limit. The hope was that this could potentially explain the acceleration of the universe \cite{eric}.

It is well-known that, unlike TEGR, generalized theories such as $f(T)$ are \emph{not} locally Lorentz invariant \cite{barrow}. 
Of course, at a purely mathematical level, a given manifold that can be parallelized admits infinitely many choices of vierbein. However, in a general teleparallel theory, such as $f(T)$ gravity, there exists a \emph{preferred frame} compatible with the field equations. [This is the difference between \emph{kinematics} and \emph{dynamics}; the latter is determined by a Lagrangian.] While TEGR has local Lorentz symmetry, and like GR has just 2 dynamical degrees of freedom,  for all non-trivial $f$'s it was thought that $f(T)$ theory would be a preferred frame theory with 5 degrees of freedom and  no local Lorentz symmetry.
The main message in a recent work by Ferraro and Fiorini \cite{1412.3424} is that this common belief is in fact much more subtle. They argued that, depending on the spacetime manifold, $f(T)$ gravity may ``inherit'' some ``remnant symmetry'' from the full [orthochronous] Lorentz group, and therefore there could exist more than one such preferred frame -- even infinitely many. 

More precisely, Ferraro and Fiorini discovered a remarkable yet simple result that $f(T)$ gravity is only invariant under Lorentz transformations of the vierbein satisfying 
\begin{equation}\label{1}
d(\epsilon_{abcd} e^a \wedge e^b \wedge \eta^{de} L^c_{~f} (L^{-1})^f_{~e,\mu} dx^\mu)=0.
\end{equation}
In other words, while this is fulfilled by a global Lorentz transformations, there are \emph{local} Lorentz transformations that could also satisfy Eq.(\ref{1}).  
The set of those local Lorentz transformations that satisfy Eq.(\ref{1}), given a frame $e_a (x)$ [or equivalently the co-frame $e^a(x)$] that solves the field equations of $f(T)$ gravity, is denoted by $\mathcal{A}(e^a)$, and dubbed the ``remnant group'' [which can be, but is not necessarily, a group]. We will refer to the additional symmetry embodied in the remnant group [in addition to the global Lorentz transformation] as the ``remnant symmetry''.

Ferraro and Fiorini then asserted that the existence of the remnant group seems to be not consistent with\footnote{Indeed, they used a stronger expression ``seems to discredit''.} the results obtained by us in \cite{1303.0993} and \cite{1309.6461}, in which we showed that $f(T)$ gravity is generically problematic -- it allows superluminal propagation and \emph{local} acausality [that is, temporal ordering is not well-defined even in an infinitesimal neighborhood]. The theory also suffers from non-unique time evolution, i.e., Cauchy problem is ill-defined. That is, given a full set of the Cauchy data, one cannot predict what will happen in the future with certainty. Note that all these problems arise at the classical level.  Ferraro and Fiorini did not explain the reason they think their results contradict ours. 

In this work we wish to clarify that the existence of the remnant group does not, in fact, contradict our previous works, but instead strongly supports it. 

\section{Comments on Propagation and Evolution in $f(T)$ Gravity}

A good way to understand the number of dynamical degrees of freedom is from the Hamiltonian perspective.
Following Dirac's procedure we find primary constraints, introduce them into the Hamiltonian with Lagrange multipliers, determine the multipliers if possible and find any additional secondary constraints.  The constraints are divided into two classes:  \emph{first class} are associated with gauge freedom, and \emph{second class} related to non-dynamical variables.
Here we are concerned with teleparallel theories.  The primary dynamic variable is the orthonormal frame $e^a{}_\mu$.  Its conjugate momentum is $P_a{}^\mu$. For the Hamiltonian analysis of teleparallel theories, see e.g., \cite{CCN, telham, BtelepHam, Bgauge}.

The Lagrangian \emph{never} contains the time derivative of $e^a{}_0$, consequently we always have the primary constraints $P_a{}^0$.  Preservation of these constraints lead to a set of 4 secondary constraints referred to as the Hamiltonian and momentum constraints.  These 8 constraints are all first class, geometrically they generate spacetime diffeomorphisms.  We need not consider them any further.
 
For the special subclass of theories of the form $f(T)$ there are 6 more primary constraints, $P^{[\mu\nu]}\simeq0$ associated with the Lorentz sector.  To keep the discussion simple and focus on the essentials, let us assume that they do not give rise to any additional constraints.  The class of these 6 constraints is determined by the rank of the associated Poisson bracket matrix.  If the rank is 0, then all constraints are first class, and they generate local Lorentz transformations --- this is the TEGR special case.  If the rank is maximal, i.e. 6, then all constraints are second class. Naturally, one would like to ask if there are other possibilities.
 
Since the second class constraints come in pairs, we might imagine we could have rank 4 [i.e., 2 extra first class and 4 second class constraints] or rank 2 [i.e., 4 extra first class and 2 second class constraints].  The latter case seems to be actually not possible.  These ``extra'' first class constraints each generates a one parameter subgroup of the Lorentz group.  If we have 2 such generators their commutator is also a generator.  Thus we would have an additional unwanted constraint, unless it vanished identically.  Therefore, by counting, a local symmetry group which is an Abelian subgroup of the Lorentz group is sensible.  It seems that there are no suitable 4 parameter subgroups of the Lorentz group.

Thus the possible scenarios seem to be (i) all the 6 constraints are second class, (ii) 2 commuting first class constraints $+$ 4 second class constraints, (iii) all constraints are first class. The standard formula for counting degrees of freedom is
\begin{flalign}
\#(\text{dof}) =&[2\cdot {\#}(\text{dynamical variables}) - 2\cdot {\#}(\text{1st class}) \notag \\ &- {\#}(\text{2nd class})]/2,
\end{flalign}
where $\#$ denotes ``the number of''. This gives for the possibilities (i,ii,iii) above as, respectively,
\begin{itemize}
\item[(i)] $[2(16)-2(8)-6]/2 = 5$; 
\item[(ii)] $[2(16)-2(10)-4]/2 = 4$;
\item[(iii)] $[2(16)-2(14)-0]/2 = 2$.
\end{itemize}

The possibility (iii) corresponds to TEGR, while (i) is the generic case for $f(T)$ gravity, as confirmed by  the detailed work of Miao Li et al. \cite{li}, which was based on the Hamiltonian analysis of \cite{maluf}. In other words, $f(T)$ gravity generically propagates 5 degrees of freedom, i.e., there are 3 additional degrees of freedom compared to standard GR. These extra degrees of freedom are very non-linear in nature. For example, they do not show up in the linear perturbation of flat Friedmann-Lema\^{i}tre-Robertson-Walker [FLRW] cosmological background \cite{1212.5774}. 

What about the option (ii) then? 
Case (ii) is ``exotic''; it is some kind of geometry that is intermediate between a metric and preferred frame theory.  Our Hamiltonian analysis identifies this possibility, which as we shall see below, \emph{is precisely one of the types that was found by Ferraro and Fiorini}.  This is the reason why the word ``generically'' is crucial, for in \cite{1303.0993} we found out that the number of physical degrees of freedom and the classes of Dirac constraints, can and do change depending on the values of the fields. That is, they are \emph{expected to be different} on different background geometries\footnote{Note that the determinant of the Poisson bracket matrix is a polynomial in the variables and their derivatives.  Generically it has real roots.  The rank cannot be constant if it admits generic solutions [which include, but is not limited to, Minkowski and FLRW].}. This is in fact, \emph{in agreement} with the findings of Ferraro and Fiorini \cite{1412.3424} that different geometries give rise to a different number of ``admissible frames''. We will further elaborate on this later.

However, \emph{precisely} because of the possibility that field configurations can change the number of degrees of freedom as well as the constraint structure of the theory,  we expect anomalous propagation such as superluminal shock waves to arise. This is explained in detail already in \cite{1303.0993} in which we employed the well-understood PDE method of characteristics, pioneered by Cauchy and Kovalevskaya. [See also Section 2 of \cite{1410.2289} and the references therein for further explanation regarding this method]. 

One has to be careful in distinguishing between the symmetry of a \emph{theory} and the symmetry of a \emph{particular solution}. For example, consider a complex scalar field $\varphi$ with a simple potential:
\begin{equation}
V = V_1(\varphi \varphi^*)  + (1-\varphi \varphi^*) V_2 (\varphi).
\end{equation}
Clearly only the $V_1$ term in the potential has local U(1) symmetry because it is a function of  $\varphi \varphi^*$, i.e., the [square of the] absolute value of $\varphi$.  On the other hand, $V_2$ term does not because it depends on the explicit $\varphi$ configuration. However, if the absolute value of $\varphi$ is unity, i.e. if $\varphi \varphi^*=1$, then the value of the potential $V$ is invariant under U(1) transformation. Thus, for some specific field configurations $V$ has U(1) symmetry, \emph{but this is not the symmetry of the theory} for generic values of $\varphi$! 
The specific values that ``restores'' U(1) symmetry to $V$ has physical effect, it is the signal that the mode related to U(1) direction has become massless. Similarly in $f(T)$ gravity one expects that if the fields evolve such that some extra symmetries emerge, there would be physical effects that accompany the changes in the number and type of constraints [much of the effort in \cite{1303.0993} and \cite{1309.6461} is spent on showing that the superluminal propagations are not simply due to gauge choice]. To be more specific, one can consider a kinetic term for $\varphi$ of the form:
\begin{equation}
(1-\varphi \varphi^*) \partial_\mu \varphi \partial^\mu \varphi^*.
\end{equation}
This kinetic term does not have \emph{local} $U(1)$ symmetry because 
the derivative is not covariant.
Generically this term is well-defined, but when the field configurations approach the value such that $\varphi \varphi^*=1$, the dynamical term vanishes. 
Under the local $U(1)$ transformation, $\varphi \varphi^*=1$ still holds and the dynamical term remains naught. 
This is indeed a ``remnant symmetry". 
However, the differential equations [the equations of motion] then behave very differently depending of the values of the fields.
This is the situation faced by $f(T)$ gravity.

In fact, the results of \cite{1412.3424} actually \emph{support} our results. As we recall, generically $f(T)$ gravity has 5 degrees of freedom. However, for almost any $f$ it is very likely that there would exist solutions where the Poisson bracket matrix has less than the generic rank.  For any such solution one or more of the generically second class constraints will now be first class.  These first class constraints will generate some local Lorentz transformations.  Thus what we expect to see is  that for each solution there is some subgroup of the Lorentz group which acts as a local symmetry gauge group. In other words, our analysis involving rank changes already implicitly implied the same result that Ferraro and Fiorini now discovered, using a different, more straightforward analysis. 

The point is that the set $\mathcal{A}(e^a)$ [which is generally not a group] encodes all information about the change in rank of the Poisson bracket matrix -- the size of this ``group" [at each spacetime point] reflects the number of normally second class constraints, which have become first class for a particular frame, i.e., the change in the rank of the Poisson bracket matrix. For a given function $f(T)$, if $\mathcal{A}(e^a)$ is empty for every solution to the equations, then the propagation has no problems. For a given $f$, and for all solution frames, if this set is an Abelian group of size \emph{independent of spacetime point}, then we may have good propagating modes. However, 
if this set is an Abelian group with size varying from point to point,
then we have acausal propagation -- which as we argued in \cite{1303.0993}, arise as a consequences of the change in the number of degrees of freedom and the constraint structure of the theory. 

Most crucially, it should be noted that in \cite{1309.6461}, we constructed an \emph{explicit} example in which $f(T)$ gravity and its Brans-Dicke-generalization with scalar field suffers from non-unique evolution -- starting with a perfectly homogeneous and isotropic, flat FLRW universe, anisotropy can suddenly emerge. 
By non-unique evolution, we do \emph{not} simply mean the following: Given a spacetime geometry, a chosen tetrad field can evolve into another choice of tetrad, which corresponds to the same metric tensor [if there are multitudes of ``admissible frames'' this would not be a problem notwithstanding the discussion above, \emph{if physical observables only couple to the metric}]. Instead we mean a stronger statement: even a \emph{geometry}, described by the metric tensor, can change drastically under the evolution, and such change cannot be predicted from initial data alone. This problem cannot be evaded even if there are more than one ``admissible frames'' corresponding to a \emph{fixed} geometry.

\section{Discussion}

In this work we discuss why the existence of remnant symmetry as shown by Ferraro and Fiorini does not contradict our previous works, which established the existence of superluminal propagation, local acausality, and non-unique time evolution in $f(T)$ gravity [and its Brans-Dicke generalization]. In fact, these problems are closely related to the remnant symmetry -- while our Hamiltonian analysis agrees with the results of Ferraro and Fiorini, the same analysis also shows that there are serious dynamical difficulties, especially if one approaches a point where the rank of the Poisson bracket matrix changes.

We now conclude this work with some additional comments.

In relation to the Hamiltonian analysis, the usual understanding is that generically teleparallel Lagrangians have no local frame gauge freedom. Beyond the relations that follow from diffeomorphism invariance [which are connected with energy momentum] the equations satisfy no other differential identities. According to Noether's Second Theorem, a local gauge freedom means a differential identity. In the Hamiltonian formulation of teleparallel theories, generically the only first class constraints are the Hamiltonian and momentum constraints. If there are no other first class constraints, then there is no other local gauge symmetries. It would be very interesting to see how ``remnant symmetries'' fit into this scheme more explicitly.

We now remark on the comment in \cite{1412.3424} on the possibility of constructing local inertial frames in $f(T)$ gravity, and the hope that Zeeman's theorem on $\Bbb{R}^{3,1}$\cite{zeeman} would thus ensure local causality. In the context of Zeeman's theorem, the invariance group $G$ of the Minkowski spacetime [the orthochronous Lorentz group, the translation group and the dilatation group] induces the light cone structure of $\Bbb{R}^{3,1}$. This structure provides some causality relation $C$ which allows the definition of causality group $G_c$. In 4 dimensions, it turns out that $G=G_c$. Indeed one could generalize the notion of Riemann normal coordinates to other geometries defined by different connections, in particular the Weitzenb\"ock connection. This was accomplished in \cite{nester}. However, propagation is \emph{not} defined at a point; it involves dynamical modes moving from a spacetime point $x$ to another [although this distance could be in a $\varepsilon$-neighborhood of $x$], and this is problematic in $f(T)$ gravity if the ``size'' of $\mathcal{A}(e^a)$ [and thus the remnant symmetry] changes from point to point. Furthermore, even local causality is certainly not avoided. To see this, one should consider also the equation that governs the propagation, derived in \cite{1303.0993}:
\begin{equation}\label{chareqn}
\left[f_T M_a^{~\mu\nu}{}_b^{~\alpha\beta}+ 2f_{TT}S_a^{~\mu\nu}S_b^{~\alpha\beta}\right]k_\mu k_\alpha \bar{e}^b_{~\beta} = 0; 
\end{equation}
in which
\begin{equation}
 M_a^{~\mu\nu}{}_b^{~\alpha\beta}:=\dfrac{\partial S_a^{~\mu\nu}}{\partial T^b_{~\alpha\beta}}.\\
\end{equation}
Here the notation $\bar{e}^b_{~\beta}$ represents the change of the frame in a certain direction, instead of the value of the frame. $k^\mu$ denotes the normal to a characteristic hypersurface which in this case could be timelike [signaling a tachyonic propagation]. Here, the first term in the square bracket is the healthy propagation; it is the only term presents in the case of TEGR, in which case $f(T)=T \Rightarrow f_{TT} \equiv 0$. This describes the normal light cones in [TE]GR. Differential equations are of course \emph{local} in nature. Thus we see that even locally the characteristic cones in a generic $f(T)$ theory is \emph{not} the same as TEGR, in particular the cones depend also on the second term [in the square bracket], which generically differs from one spacetime point to another. 

Lastly we would like to make a side remark that in \cite{1412.3424}, two classic works by Hayashi and Shirafuji were mentioned \cite{HS1, HS2}. In particular, it was pointed out that \cite{HS1} considered ``restricted local invariance of this sort'' in the context of ``New General Relativity'' [NGR] proposed in \cite{HS2}. 
Indeed Hayashi and Shirafuji considered the possibility of a new type of geometry somewhere in between teleparallel and Riemannian one.  It would have what we have here called ``remnant symmetry'', a preferred frame determined up to a certain dynamically determined subgroup of the Lorentz group.   
It is worth mentioning that there has since been some criticisms by Kopczy\'nski and further developments that came out of that \cite{K, N, CCN, CNY}. [The work by Chen, Nester and Yo \cite{CNY} was the first to call attention to the effect of non-linear constraints and the relationship of changes in the rank of the Poisson matrix with tachyonic characteristics]. Kopczy\'nski's objection was in regard to one of the complications that
occurs when one has some ``local symmetry" for the gravitational Lagrangian for a certain subclass of solutions. Briefly, if the LHS of the field
equation has a local symmetry, then the RHS, i.e., the material energy
momentum tensor must have the same symmetry. This could impose
``unphysical" limitations on the matter sector. For the NGR theory, it was found that spin-1/2 Dirac field does not give rise to any problem, but for a hypothetical spin-3/2 field there is inconsistency [unless if one subscribes to non-minimal coupling to save the theory].  For $f(T)$ gravity this may not be a serious problem, as one could simply take the source energy-momentum tensor to be completely locally Lorentz invariant, as it is in GR. However this would mean that there is no material source for the extra degrees of freedom.

Despite the pathologies of $f(T)$ gravity, the existence of remnant symmetry is indeed interesting and may provide further insights into the structure of teleparallel gravities, and perhaps modified gravity theories in general.   

\begin{acknowledgments}
The authors would like to thank Huan-Hsin Tseng for discussion. 
K.I. is supported by Taiwan National Science Council under Project No. NSC101-2811-M-002-103.
\end{acknowledgments}


\end{document}